\documentstyle[twoside,fleqn,espcrc2,psfig]{article}
\begin{document}
\title{Abelian dominance in chiral symmetry breaking
\thanks{presented by F. X. Lee}}
\author{Frank X. Lee\address{TRIUMF, 4004 Wesbrook Mall,
Vancouver, British Columbia, Canada V6T 2A3},
R. M. Woloshyn$^a$
and Howard D. Trottier\address{Department of Physics, 
Simon Fraser University, Burnaby, British Columbia, Canada V5A 1S6}}

\begin{abstract}
Calculations of the chiral condensate
$\langle \bar{\psi} \psi \rangle$ on the lattice  
using staggered fermions and the Lanczos algorithm
are presented.
Three gauge fields are considered: 
the quenched non-Abelian field,
 the Abelian field
projected in the maximal Abelian gauge, and the monopole
field further decomposed from the Abelian field. 
The results show that the Abelian monopoles 
largely reproduce the chiral condensate values of 
the full non-Abelian theory, both in SU(2) and in SU(3).
\end{abstract}
\maketitle

\section{Introduction}

Since the Abelian monopole mechanism for
confinement in QCD was conjectured~\cite{hoof1,hoof2},
there have been extensive studies using lattice gauge 
theory in the pure gauge 
sector~\cite{suzuki,kron,kron1,brand,deb,yee,poulis}.   
Only a small amount of work has been done in the
quark sector~\cite{rmw,miya,suzuki1,miya1}.

In the present work, we do a systematic study of the 
role of Abelian projection and 
Abelian monopoles in 
chiral symmetry breaking in the confined phase at zero temperature. 
We use a different approach from that used in~\cite{rmw},
namely, the eigenvalue method using the Lanczos algorithm~\cite{lanc},
which allows us to work at zero mass.
In addition to working in SU(2) gauge theory, 
we also perform calculations in SU(3) for which there has been very little
study of the Abelian monopole mechanism.

\section{Chiral Condensate $\langle \bar{\psi} \psi \rangle$ on the lattice}

The spontaneous breakdown of chiral symmetry 
is signaled by the
non-vanishing of the order parameter 
(chiral condensate) $\langle \bar{\psi} \psi \rangle$. 
On a lattice of volume V, it is given by
\begin{equation}
\langle \bar{\psi} \psi(m,V) \rangle = -{1\over V} \langle
\mbox{Tr} \left[ {1\over D\hspace{-2.5mm}/ + m} \right] \rangle
\end{equation}
where the angular brakets denote the gauge field configuration average.
This can be expressed in terms of the eigenvalues of the
Dirac operator $D\hspace{-2.5mm}/$ 
\begin{eqnarray}
\langle \bar{\chi} \chi(m,V) \rangle &=&{-1\over V}
\sum_{n=1}^N {1\over i\lambda_n +m} 
\nonumber \\ 
&=&{-1\over V}\sum_{\lambda_n\geq 0}{2m\over \lambda_n^2 +m^2}.
\end{eqnarray}
Here $\bar{\chi}$, $\chi$ are the single-component staggered fermion
fields and $\langle \bar{\psi} \psi \rangle=\frac{1}{4}\langle
\bar{\chi} \chi \rangle$. The eigenvalues are calculated using
the well-established Lanczos algorithm~\cite{lanc}.
To truly probe the physics of spontaneous chiral symmetry breaking, 
one should attempt to work in the limit of zero quark mass and 
infinite volume. The chiral limit $m\rightarrow 0$  should be taken 
after $V\rightarrow \infty$
\begin{eqnarray}
\langle \bar{\chi} \chi \rangle &=&
- \lim_{m\rightarrow 0} \lim_{V\rightarrow \infty}
{1\over V}\sum_{\lambda_n\geq 0}{2m\over \lambda_n^2 +m^2} 
\nonumber \\ 
&=& -\lim_{m\rightarrow 0} \int_0^\infty
  d\lambda{2m \rho(\lambda) \over \lambda^2 +m^2}
=-\pi\rho(0)
\label{rho}
\end{eqnarray}
where $\rho(\lambda)={1\over V}{dn/ d\lambda}$ is 
called the spectral density function and 
is normalized to $\int_0^\infty d\lambda\rho(\lambda)=N_c$, the 
number of colors. Eq.~(\ref{rho}) 
relates chiral symmetry breaking to the small modes in 
the  eigenvalue spectrum.
So the task is reduced to finding the small eigenvalues, 
rather than the entire spectrum of the fermion matrix.

\section{Abelian Projection on the Lattice} 

The lattice formulation of Abelian projection was developed 
in~\cite{kron,kron1}. The idea is to fix the gauge of a SU(N) theory 
so that 
a residual gauge freedom $U(1)^{N-1}$ remains. The Abelian degrees of 
freedom are extracted by a subsequent projection (Abelian projection):
$ U(x,\mu)=c(x,\mu)\;u(x,\mu)$
where $u(x,\mu)$ is the diagonal {\em abelian-projected field} and 
$c(x,\mu)$ the nondiagonal matter (gluon) field. 
In general the gauge condition can be 
realized by making some adjoint operator ${\cal R}$
diagonal: 
\begin{equation}
\tilde{\cal R}(x)=G(x){\cal R}(x)G^{-1}(x)=\mbox{diagonal}.
\label{diag}
\end{equation}

Several gauge conditions have been studied
and it has been found that the so-called Maximal Abelian gauge~\cite{kron1}
most readily captures the long-distance features of the vacuum.
It is realized by maximizing the quantity, in SU(2),
\begin{equation}
 R=\sum_{x,\mu} \mbox{tr} \left[
\sigma_3 \tilde{U}(x,\mu)\sigma_3 \tilde{U}^\dagger(x,\mu)\right]
\end{equation}
or in SU(3)
\begin{equation}
R=\sum_{x,\mu}\sum_{i=1}^{3} |\tilde{U}_{ii}(x,\mu)|^2
\end{equation}
where $\tilde{U}(x,\mu)=G(x)U(x,\mu)G^{-1}(x+\mu)$.
In practice, $R$ is maximized iteratively by solving 
for $G(x)$ repeatedly until some criterion is satisfied. 
It is well-known that there exist monopoles in a compact 
U(1) field as topological fluctuations.
The Abelian-projected field $u(x,\mu)$ can be further decomposed 
into monopole plus photon contributions~\cite{smit}.

\section{Numerical results}

\subsection{Results for SU(2)}

 \begin{figure*}
 \psfig{file=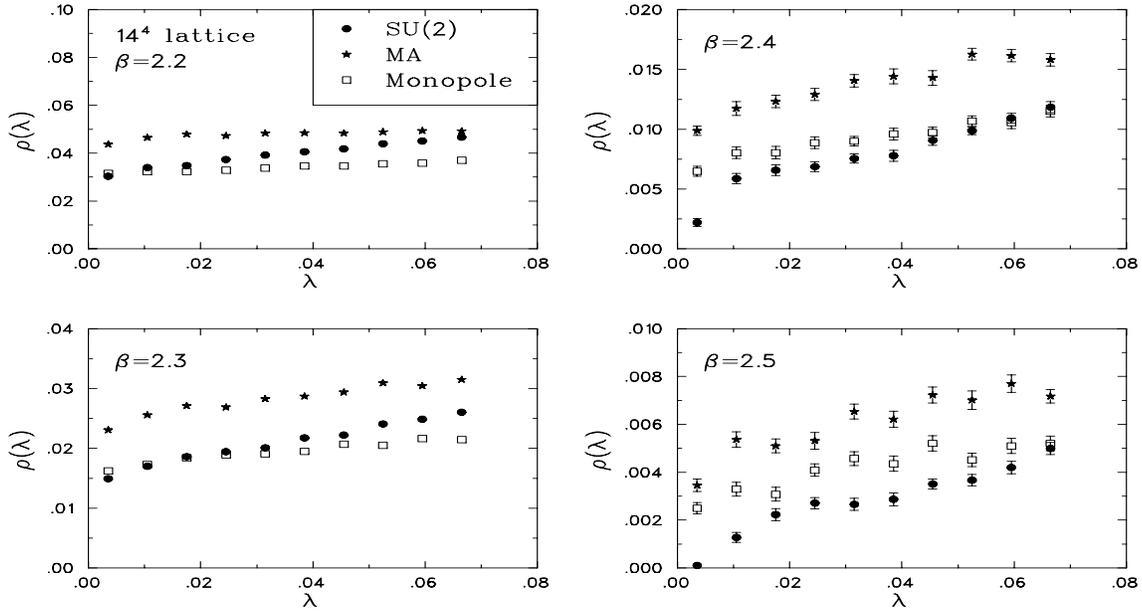,height=8cm,width=15cm,angle=90} 
 \caption{Raw data for the spectral density functions in SU(2).} 
 \label{raw2}
 \end{figure*}
\begin{center}
 \begin{figure*}
\psfig{file=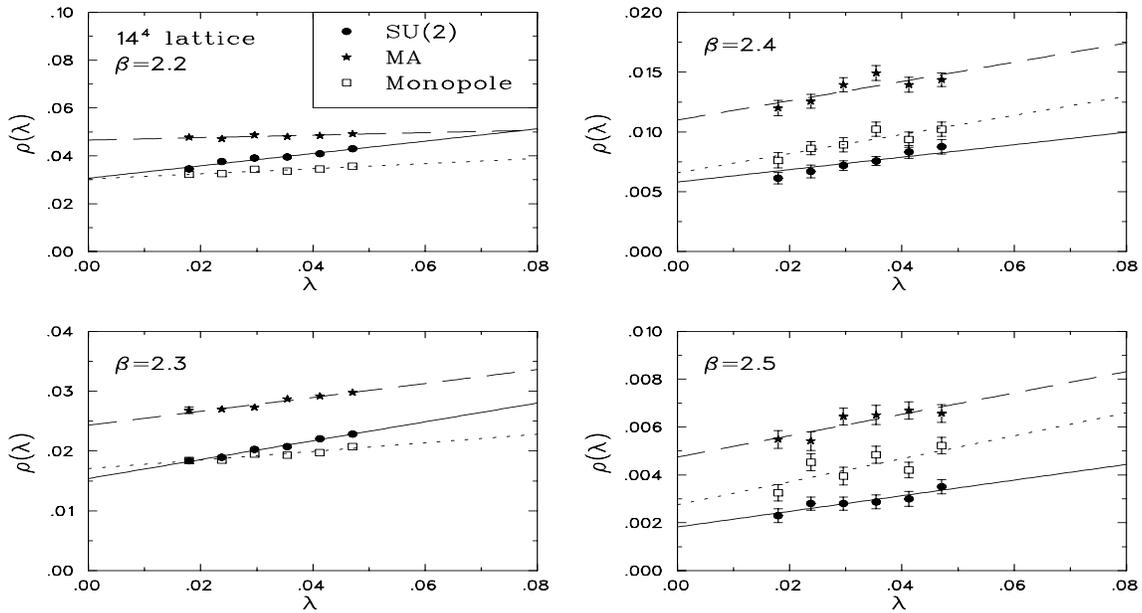,height=8cm,width=15cm,angle=90} 
 \caption{Fitted spectral density functions in the eigenvalue 
interval [0.015,0.05] in SU(2).}
 \label{fit2}
 \end{figure*}
\end{center}
Gauge fixing was done with the help of
overrelaxation~\cite{man} which reduced the number of iterations by 
a factor of 3 to 5. We used the stopping criterion that 
$\delta=1-1/2\mbox{Tr}G(x)=1-r_0$ converges to $10^{-6}$ which 
requires about 500 iterations with overrelaxation. 
Fig.~\ref{raw2} shows the raw data obtained for $\rho(\lambda)$.
To extract a value at $\lambda=0$, we fit the distributions by 
a straight line $\rho(\lambda)=a+b\lambda$ in an interval 
$[\lambda_{min},\lambda_{max}]$ which excludes eigenvalues 
strongly influenced by finite volume effects~\cite{hands}. 
Fig.~\ref{fit2} shows the results of such a 
fit for the interval [0.015,0.05]. 
The full and Abelian values for the extracted 
$\langle \bar{\chi} \chi \rangle$
are consistent with those obtained 
in Ref.~\cite{rmw} which used a quite different approach. 
The interesting feature here is that using the monopole contribution 
brings the values even closer 
to those of the full theory, while the photon configurations 
give almost no effects. 
We found either no or very few small eigenvalues for the photon
configurations.
For purposes of comparison, we also performed Abelian projection
at $\beta=2.5$ using a different gauge-fixing condition: the 
Polyakov gauge which diagonalizes the Polyakov loop according to 
Eq.~(\ref{diag}). The result is that the Abelian and the 
monopole spectral density functions 
are almost an order of magnitude larger than those of the full theory. 
In Ref.~\cite{rmw}, a similar result is found using 
the field-strength gauge.

\subsection{Results for SU(3)}

\begin{figure}[t]
\psfig{file=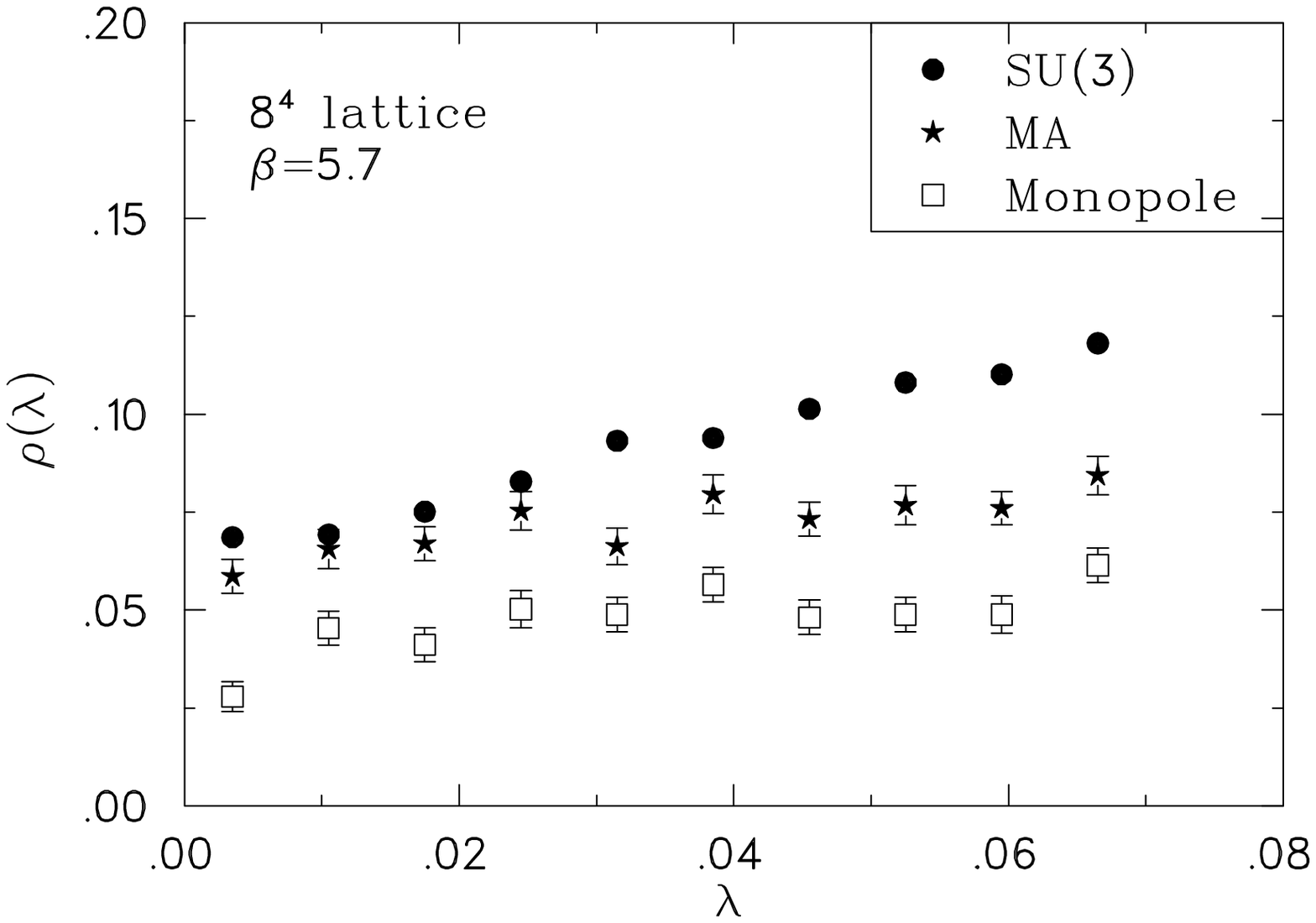,height=5.5cm,width=7.5cm,angle=90}
 \caption{Spectral density function in SU(3) on the
$8^4$ lattice at $\beta$=5.7.}
 \label{ch8b57}
 \end{figure}

\begin{figure}
\psfig{file=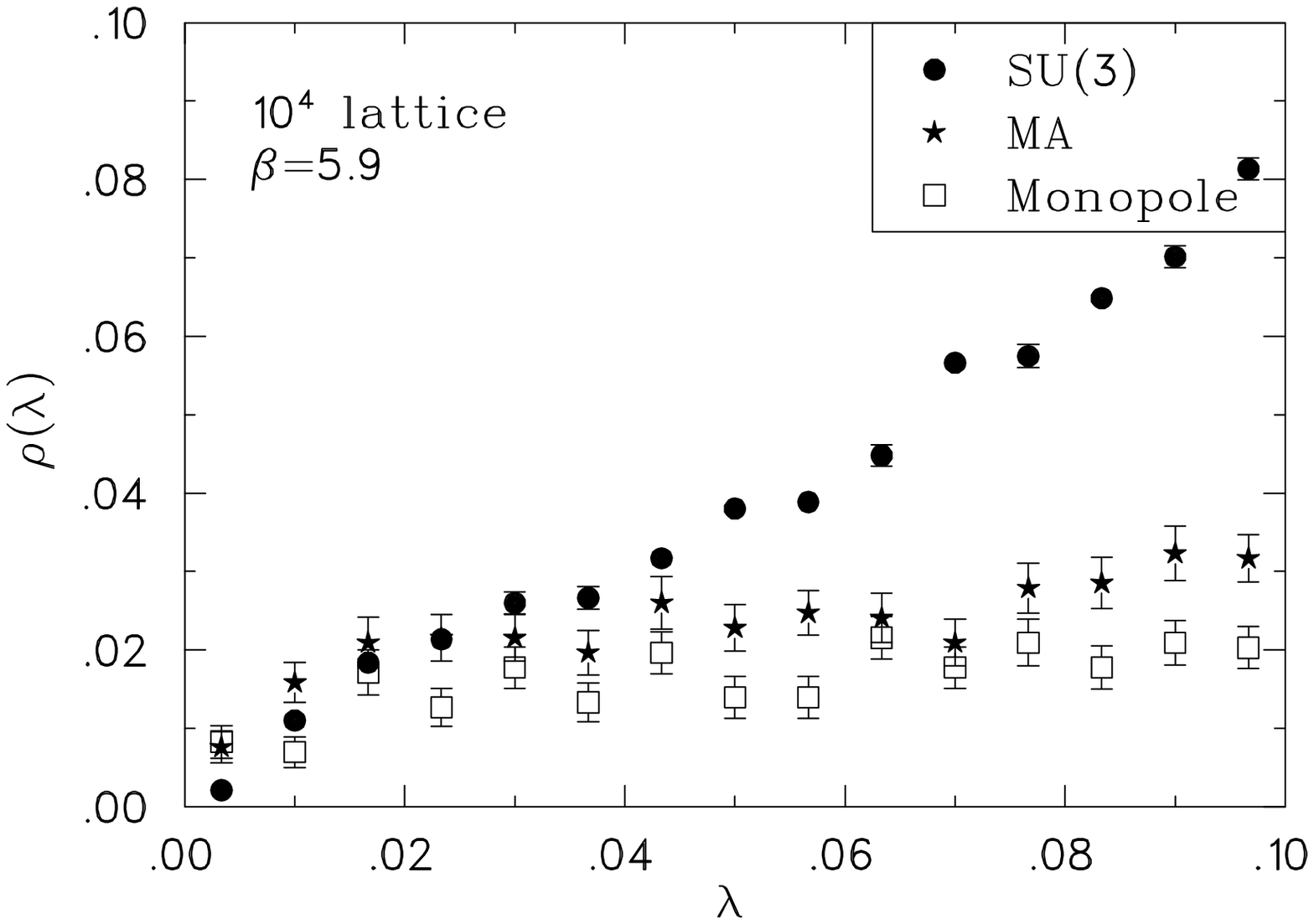,height=5.5cm,width=7.5cm,angle=90}
 \caption{Spectral density function in SU(3) on the
$10^4$ lattice at $\beta$=5.9.}
 \label{ch10b59}
 \end{figure}

The gauge field configurations were generated using the 
Cabibbo-Marinari~\cite{cab} pseudo-heat-bath method on a 
$8^4$ lattice at $\beta=5.7$ and a $10^4$ lattice at $\beta=5.9$.
Configurations are selected after 4000 thermalization sweeps from a 
cold start, and every 500 sweeps thereafter.
Fig.~\ref{ch8b57} shows the results obtained
for 150 configurations on the $8^4$ lattice.
Fig.~\ref{ch10b59} shows the results obtained
for 100 configurations on the $10^4$ lattice.
One can see that a similar pattern emerges in SU(3):
the Abelian and the monopole 
contributions give values that are close to those of the full theory.
It was also confirmed that the photon configurations give 
negligible contribution.

\section{Conclusion}
We have calculated chiral symmetry breaking on the lattice in
the quenched approximation using the Lanczos algorithm which allows
calculations to be done directly at zero quark mass.
The results show that Abelian projected fields and
Abelian monopoles in the Maximal Abelian gauge can largely 
reproduce the values of the full theory, both in SU(2) and in SU(3).
These results extend the idea of Abelian dominance 
and provide some evidence that Abelian monopoles
can describe the long-distance physics of light quarks in QCD.

\section{Acknowledgments}
This work was supported in part by the Natural Sciences and Engineering
Council of Canada.

\end{document}